\documentclass[conference]{IEEEtran}
\IEEEoverridecommandlockouts
\usepackage{cite}
\usepackage{amsmath,amssymb,amsfonts}
\usepackage{algorithmic}
\usepackage{graphicx}
\usepackage{textcomp}
\usepackage{xcolor}
\usepackage{float}
\usepackage{amsthm}
\usepackage{graphicx}
\usepackage{epstopdf}
\usepackage{amsmath,bm}
\usepackage{amsfonts}
\usepackage{amssymb}
\usepackage{color}
\usepackage{multirow}
\usepackage{multicol}
\usepackage{soul,xcolor}

\definecolor{orange}{RGB}{0,112,192}
\def\BibTeX{{\rm B\kern-.05em{\sc i\kern-.025em b}\kern-.08em
    T\kern-.1667em\lower.7ex\hbox{E}\kern-.125emX}}
\begin{document}
	
\setstcolor{red}

\makeatletter
\newcommand*{\rom}[1]{\expandafter\@slowromancap\romannumeral #1@}
\makeatother
\title{Channel Estimation under Hardware Impairments: Bayesian Methods versus Deep Learning
\thanks{The authors were supported by ELLIIT and the Wallenberg AI, Autonomous Systems and Software Program (WASP).}
}

\author{\IEEEauthorblockN{\"Ozlem Tugfe Demir and Emil Bj\"ornson}
\IEEEauthorblockA{{Department of Electrical Engineering (ISY), Link\"oping University, Sweden
} \\
{Email: \{ozlem.tugfe.demir, emil.bjornson\}@liu.se}
}

}

\maketitle

\begin{abstract}
This paper considers the impact of general hardware impairments in a multiple-antenna base station and user equipments on the uplink performance. First, the effective channels are analytically derived for distortion-aware receivers when using finite-sized signal constellations. Next, a deep feedforward neural network is designed and trained to estimate the effective channels.
Its performance is compared with state-of-the-art distortion-aware and unaware Bayesian linear minimum mean-squared error (LMMSE) estimators. 
The proposed deep learning approach improves the estimation quality by exploiting impairment characteristics, while LMMSE methods treat distortion as noise.%
\end{abstract}


\section{Introduction}

Channel estimation in multi-user multiple-input multiple-output (MIMO) is a well-studied problem \cite{Kotecha2004a,Neumann2018,emil_book}, but only in the case of ideal hardware at both the base station (BS) and user equipments (UEs). In practice, there are transceiver impairments, such as non-linearities in amplifiers, I/Q imbalance, and quantization errors \cite{book_rf}. The non-linear system behavior is often treated by utilizing the Bussgang decomposition to find an equivalent linear system with uncorrelated distortion \cite{emil_nonideal}. One can then derive a distortion-aware Bayesian LMMSE estimator that utilizes second-order distortion statistics to estimate the channels, but in doing so the distortion is treated as independent colored noise, although it depends on the channel. 

Having access to accurate channel estimates is particularly important in the presence of non-ideal hardware. Recently,\cite{emil_hardware} proposed a distortion-aware receiver for uplink signal detection in Massive MIMO. To apply this receiver, it is necessary for the BS to know the effective channels of the UEs together with the received signal correlation matrix. This has motivated us to consider the estimation of the effective channels, taking into account the BS and UE hardware distortion characteristics, instead of only the wireless channels.

The first novelty of this paper is the derivation of effective channels for symmetric finite-sized constellations in the uplink data transmission. In order to model the BS non-linearities, a quasi-memoryless polynomial model, which can represent both AM/AM and AM/PM distortions, is used in accordance with previous literature \cite{christopher,pa,ericsson}. We utilize the derived model to design a novel deep-learning-based channel estimator that we train to exploit the full structure of the hardware impairments, instead of treating the distortion as independent noise as in previous work. We compare our novel solution with both distortion-aware and unaware LMMSE estimators and show that the deep-learning-based alternative significantly outperforms them in all the considered simulation scenarios.

\section{System Model with Hardware Impairments}
We consider a single-cell multi-user MIMO system where a BS equipped with $M$ antennas serves $K$ single-antenna UEs. We focus on the uplink with non-ideal BS and UE hardware. A block-fading model is considered where the wireless channels between each BS antenna and UE is represented by a constant complex-valued scalar that takes an independent realization in each time-frequency coherence block \cite{emil_book}. In each block, the channels are estimated by sending uplink pilot sequences. 

In any arbitrary coherence block, the noise and BS-distortion-free signal ${\bf u}=[ u_1 \ \ldots \ u_M]^T \in \mathbb{C}^M$ at the input of the receive antennas in the data transmission phase is
\begin{align}
{\bf u}=\sum_{k=1}^K{\bf g}_ks_k={\bf G}{\bf s},
\end{align}
where ${\bf G}=[{\bf g}_1 \ \ldots \ {\bf g}_K] \in \mathbb{C}^{M\times K}$ is the concatenated channel matrix where ${\bf g}_k=[g_{k1}\ \ldots \ g_{kM}]^T \in \mathbb{C}^M$ is the channel from the $k^\textrm{th}$ UE to the BS. The signal from the $k^\textrm{th}$ UE is $s_k \in  \mathbb{C}$ and ${\bf s}=[s_1 \ \ldots \ s_K]^T \in \mathbb{C}^{K}$. The information signals are independent and $\mathbb{E}\{|s_k|^2\}=p_k$ for $k=1,\ldots,K$.

The channel between the $k^\textrm{th}$ UE and the $m^\textrm{th}$ BS antenna is $g_{km}=\sqrt{\beta_k}h_{km}$, $k=1,\ldots,K$, $m=1,\ldots,M,$ where $\beta_k$ is the large-scale fading coefficient, which is the same for all BS antennas in accordance with the existing literature \cite{emil_book}. The small-scale fading $h_{km} \in \mathbb{C}$ is modeled as i.i.d.~Rayleigh fading:  $h_{km}\sim \mathcal{N}_{\mathbb{C}}(0,1)$. We will now investigate the effect that non-ideal hardware has on ${\bf u}$. 

\subsection{Quasi-Memoryless Polynomial Model of BS Impairments}

The non-ideal BS receiver hardware is modeled as a non-linear quasi-memoryless function where both the amplitude and phase of the received signal are distorted. We use the following general $(2L+1)^\textrm{th}$ order polynomial model for this kind of distortion in the complex baseband \cite{book_rf}, \cite{christopher}:
\begin{align} \label{eq:z_m}
z_m=\sum_{l=0}^L\tilde{a}_{lm}u_m|u_m|^{2l}, \ \ m=1,\ldots,M,
\end{align}
where $z_m$ is the noise-free distorted signal at the $m^\textrm{th}$ BS antenna and $\{\tilde{a}_{lm}\}$ are complex scalars, which means that both AM/AM and AM/PM distortion are considered \cite{book_rf}. The model in \eqref{eq:z_m} describes the non-linearities by the joint effect of  amplifiers, local oscillators and mixers.  
We assume that long-term automatic gain control is utilized, thus $\tilde{a}_{lm}$ can be represented by
\vspace{-0.1cm}
\begin{align} &\tilde{a}_{lm}=\frac{a_{lm}}{{\big(b_{\text{off}}\mathbb{E}\{|u_m|^2\}\big)^l}}=\frac{a_{lm}}{\big(b_{\text{off}}\sum_{k=1}^K\beta_kp_k\big)^l} \nonumber \\
& \quad l=0,\ldots,L, \ \ \ m=1,\ldots,M,
\end{align}
where $\{a_{lm}\}$ are the reference polynomial coefficients when the signal has a magnitude between zero and one \cite{ericsson}. 
The parameter $b_{\text{off}}$ models the backoff that is used to prevent clipping \cite{emil_hardware}. Using \eqref{eq:z_m}, the digital baseband signal ${\bf y}=[y_1 \ \ldots \ y_M]^T \in \mathbb{C}^{M}$ at the BS is given by ${\bf y}={\bf z}+{\bf n}$ where  ${\bf z}=[ z_1 \ \ldots \ z_M]^T \in \mathbb{C}^{M}$ is the hardware-distorted signal from \eqref{eq:z_m} and ${\bf n} \sim \mathcal{N}_{\mathbb{C}}({\bf 0},\sigma^2{\bf I}_M)$ is uncorrelated noise. In practice, the initial noise entering into the BS hardware is also affected by the nonlinear distortion, however the resultant noise is still uncorrelated with ${\bf u}$. For analytical tractability, this model is used in accordance with \cite{emil_hardware}.

\vspace{-0.16cm}
\subsection{Modeling of UE Hardware Impairments}

Since impairments in the UE hardware can be the performance-limiting factor \cite{emil_book,emil_nonideal,emil_hardware}, we include this in our model. Following \cite{emil_hardware}, we assume that  $s_k=\sqrt{\kappa_kp_k}\varsigma_k+\omega_k$ for $k=1,\ldots,K$, where $\varsigma_k$ is the zero-mean data signal transmitted by the $k^\textrm{th}$ UE and $\mathbb{E}\{|\varsigma_k|^2\}=1$ and $\omega_k$ is independent circularly symmetric distortion with variance $(1-\kappa_k)p_k$.
The parameter $\kappa_k \in [0,1]$ quantifies the level of hardware impairment at the $k^\textrm{th}$ UE, after signal predistortion.

\vspace{-0.16cm}
\section{Effective Channels for Symmetric Finite-Sized Signal Constellations}
 
We will now derive the effective channel during data transmission with distortion-aware receivers, which includes the wireless channel and hardware impairments.
A similar impairment model was considered in \cite{emil_hardware} but with complex Gaussian information signals $\{\varsigma_k\}$. Different from \cite{emil_hardware}, we consider symmetric finite-sized constellations as in practice.

We consider a fixed channel realization ${\bf G}$ in an arbitrary coherence block and let $\mathbb{E}_{|{\bf G}}\{.\}$ denote the conditional expectation given ${\bf G}$. Following the approach in \cite{emil_hardware}, the digital baseband signal ${\bf y}$ can be written as a summation of the LMMSE estimate of ${\bf y}$ given $\bm{\varsigma}=[\varsigma_1 \ \ldots \ \varsigma_K]^T$ plus the additive distortion and noise term as follows:
\begin{align} \label{eq:Bussgang}
{\bf y}={\bf C}_{y\varsigma}{\bf C}_{\varsigma\varsigma}^{-1}\bm{\varsigma}+\bm{\eta}
\end{align}
where ${\bf C}_{y\varsigma}=\mathbb{E}_{|{\bf G}}\{{\bf y}\bm{\varsigma}^H\}$ and ${\bf C}_{\varsigma\varsigma}=\mathbb{E}_{|{\bf G}}\{\bm{\varsigma}\bm{\varsigma}^H\}=\mathbb{E}\{\bm{\varsigma}\bm{\varsigma}^H\}={\bf I}_K$. The additive distortion $\bm{\eta}={\bf y}-{\bf C}_{y\varsigma}{\bf C}_{\varsigma\varsigma}^{-1}\bm{\varsigma}$ is uncorrelated with $\bm{\varsigma}$ by construction. We call $ {\bf C}_{y\varsigma}$ the effective channel since the signal term in \eqref{eq:Bussgang} is ${\bf C}_{y\varsigma}{\bf C}_{\varsigma\varsigma}^{-1}\bm{\varsigma} =  {\bf C}_{y\varsigma} \bm{\varsigma} $. 
To derive the elements of ${\bf C}_{y\varsigma}$, we first define the signals
\vspace{-0.12cm}
\begin{align}
t_m=\sum_{k=1}^K\sqrt{\kappa_kp_k}g_{km}\varsigma_k, \ \  v_m=\sum_{k=1}^Kg_{km}\omega_k, \  \ m=1,\ldots,M,
\end{align}
where $u_m=t_m+v_m$ for $m=1,\ldots,M$ and $v_m$ is conditionally independent of $t_m$ with power $\sum_{k=1}^K|g_{km}|^2(1-\kappa_k)p_k$ for $m=1,\ldots,M$. The $(m,k)^\textrm{th}$ element of the effective channel matrix ${\bf C}_{y\varsigma}$, i.e., $[{\bf C}_{y\varsigma}]_{mk}$ is given by
\begin{align} \label{eq:C_ysigma}
[{\bf C}_{y\varsigma}]_{mk}&=\mathbb{E}_{|{\bf G}}\{y_m\varsigma_k^{*}\}\nonumber \\&=\sum_{l=0}^L\tilde{a}_{lm}\mathbb{E}_{|{\bf G}}\{|t_m+v_m|^{2l}(t_m+v_m)\varsigma_k^{*}\} \nonumber \\
&=\sum_{l=0}^L\tilde{a}_{lm}\sum_{\substack{l_1,l_2,l_3,l_4 \\l_1+l_2+l_3+l_4=l}}\binom{l}{l_1,l_2,l_3,l_4}\times \nonumber \\
&\mathbb{E}_{|{\bf G}}\{|t_m|^{2l_1}|v_m|^{2l_2}(t_mv_m^{*})^{l_3}(t_m^{*}v_m)^{l_4}(t_m+v_m)\varsigma_k^{*}\} \nonumber \\
&=\sum_{l=0}^L\tilde{a}_{lm}\sum_{\substack{l_1,l_2,l_3,l_4 \\l_1+l_2+l_3+l_4=l}}\binom{l}{l_1,l_2,l_3,l_4}\times\nonumber\\ &\mathbb{E}_{|{\bf G}}\{|t_m|^{2l_1}t_m^{l_3+1}{t_m^{*}}^{l_4}\varsigma_k^{*}\}\mathbb{E}_{|{\bf G}}\{|v_m|^{2l_2}{v_m^{*}}^{l_3}v_m^{l_4}\} \nonumber \\
& +\sum_{l=0}^L\tilde{a}_{lm}\sum_{\substack{l_1,l_2,l_3,l_4\\l_1+l_2+l_3+l_4=l}}\binom{l}{l_1,l_2,l_3,l_4}\times \nonumber \\& \mathbb{E}_{|{\bf G}}\{|t_m|^{2l_1}t_m^{l_3}{t_m^{*}}^{l_4}\varsigma_k^{*}\}\mathbb{E}_{|{\bf G}}\{|v_m|^{2l_2}{v_m^{*}}^{l_3}v_m^{l_4+1}\}
\end{align}
where we used the conditional independence of $v_m$ with $t_m$ and $\varsigma_k$. Note that $v_m$ is a circularly symmetric random variable given ${\bf G}$ and hence $\mathbb{E}_{|{\bf G}}\{|v_m|^{2l_2}{v_m^{*}}^{l_3}v_m^{l_4}\}=0$ for $l_3\neq l_4$ and it is equal to $\mu_{m,(l_2+l_3)}\triangleq\mathbb{E}_{|{\bf G}}\{|v_m|^{2(l_2+l_3)}\}$ when $l_3=l_4$. Using these properties, \eqref{eq:C_ysigma} can be simplified as follows:
\begin{align} \label{eq:C_ysigma_simplified}
[{\bf C}_{y\varsigma}]_{mk}&=\sum_{l=0}^L\tilde{a}_{lm}\sum_{\substack{l_1,l_2,l_3 \\l_1+l_2+2l_3=l}}\binom{l}{l_1,l_2,l_3,l_3}\mu_{m,(l_2+l_3)}\times \nonumber\\ &\mathbb{E}_{|{\bf G}}\{|t_m|^{2(l_1+l_3)}t_m\varsigma_k^{*}\} \nonumber \\
& +\sum_{l=0}^L\tilde{a}_{lm}\sum_{\substack{l_1,l_2,l_3 \\ l_1+l_2+2l_3-1=l}}\binom{l}{l_1,l_2,l_3,l_3-1}\mu_{m,(l_2+l_3)}\times \nonumber \\ & \mathbb{E}_{|{\bf G}}\{|t_m|^{2(l_1+l_3-1)}t_m\varsigma_k^{*}\}.
\end{align}
If we define $r_1=l_1+l_3$ and $r_2=l_1+l_3-1$, \eqref{eq:C_ysigma_simplified} becomes
\begin{align} \label{eq:C_ysigma_simplified2}
[{\bf C}_{y\varsigma}]_{mk}&=\sum_{r_1=0}^L\mathbb{E}_{|{\bf G}}\{|t_m|^{2r_1}t_m\varsigma_k^{*}\}\sum_{l=r_1}^L\tilde{a}_{lm}\mu_{m,(l-r_1)}\times \nonumber \\ &\sum_{\substack{l_1,l_2 \\l_2-l_1=l-2r_1}}\binom{l}{l_1,l_2,r_1-l_1,r_1-l_1} \nonumber \\
& +\sum_{r_2=0}^L\mathbb{E}_{|{\bf G}}\{|t_m|^{2r_2}t_m\varsigma_k^{*}\}\sum_{l=r_2}^L\tilde{a}_{lm}\mu_{m,(l-r_2)}\times \nonumber \\ &\sum_{\substack{l_1,l_2\\l_2-l_1=l-2r_2-1}}\binom{l}{l_1,l_2,r_2+1-l_1,r_2-l_1} \nonumber \\
&=\sum_{r=0}^L\mathbb{E}_{|{\bf G}}\{|t_m|^{2r}t_m\varsigma_k^{*}\}\sum_{l=r}^{L}\tilde{a}_{lm}\mu_{m,(l-r)}c_{lr}
\end{align}
where $c_{lr}$ is defined as $c_{lr}\triangleq\sum_{\substack{l_1,l_2\\l_2-l_1=l-2r}}\binom{l}{l_1,l_2,r-l_1,r-l_1}+\sum_{\substack{l_1,l_2\\l_2-l_1=l-2r-1}}\binom{l}{l_1,l_2,r+1-l_1,r-l_1}$ for $r=0,\ldots,L$ and $l=r,r+1,\ldots,L$. Note that $\{c_{lr}\}$ are independent of the channel realizations and the coefficients $\{\tilde{a}_{lm}\}$ that characterize the hardware. Hence, it is enough to calculate them offline and then use them for calculation of \eqref{eq:C_ysigma_simplified2}. Now, we only have to calculate the conditional moments $\mathbb{E}_{|{\bf G}}\{|t_m|^{2r}t_m\varsigma_k^{*}\}$ for $r=0,\ldots,L$, $m=1,\ldots,M$, and $k=1,\ldots,K$. We assume standard finite-sized constellations that satisfy the $90^{\circ}$ circular shift symmetry. This kind of symmetry implies that if $\varsigma$ is a point in the constellation, then $\varsigma e^{j\frac{\pi}{2}s}$ for $s=1,2,3$ is also a constellation point. This kind of symmetry exists in most practically used constellations: PSK of dimension divisible by four, square QAM, circular QAM, etc. For these constellations, it is easy to prove that  $\mathbb{E}\{\varsigma_k^{l_1}{\varsigma_k^{*}}^{l_2}\}=0$ if $l_1-l_2\neq 4i$ for any integer $i$ under the equal symbol probability assumption. In addition, let us define $\tilde{g}_{km}=\sqrt{\kappa_kp_k}g_{km}$ for $m=1,\ldots,M$ and $k=1,\ldots,K$ and $S_r=\min(r+1,K)$.  Using these properties,  $\mathbb{E}_{|{\bf G}}\{|t_m|^{2r}t_m\varsigma_k^{*}\}$ can be expressed as follows:
\begin{align} \label{eq:moments}
&\mathbb{E}_{|{\bf G}}\{|t_m|^{2r}t_m\varsigma_k^{*}\} \nonumber \\&=\mathbb{E}_{|{\bf G}}\bigg\{\bigg(  \sum_{l=1}^K\tilde{g}_{lm}\varsigma_l\bigg)^{r+1}\bigg(  \sum_{l=1}^K\tilde{g}_{lm}^{*}\varsigma_l^{*}\bigg)^r\varsigma_k^{*}\bigg\} \nonumber \\
&=\sum_{\substack{k_1,\ldots,k_{S_r}, l_1,\ldots,l_{S_r}\\k_1+k_2+\ldots+k_{S_r}=r+1,\\l_1+l_2+\ldots+l_{S_r}=r+1, \\ k_s-l_s=4i \text{ for some integer} \ i \text{ for } s=1,\ldots,S_r, \\ l_1\geq 1, \\ k_{S_r}\geq k_{S_r-1} \geq \ldots \geq k_2, \\ l_s\geq l_{s-1} \text{ if } k_s=k_{s-1} \text{ for } s=3,\ldots,S_r}}  \binom{r+1}{k_1,k_2,\ldots,k_{S_r}} \times \nonumber \\ &\binom{r}{l_1-1,l_2,\ldots,l_{S_r}}  \prod_{s=1}^{S_r}\mathbb{E}\{\varsigma^{k_s}{\varsigma^{*}}^{l_s}\}{\tilde{g}_{km}}^{k_1}({\tilde{g}_{km}}^{*})^{l_1-1} \times \nonumber \\ & \sum_{\substack{f_2,\ldots,f_{S_r}\\ f_i\neq f_j \text{ for } i\neq j\\f_i\neq k \text{ for } i=2,\ldots,S_r\\ f_i>f_j \text{ if } k_i=k_j \text{ and } l_i=l_j \text{ for } i\neq j} }\prod_{i=2}^{S_r}\tilde{g}_{f_im}^{k_i}({\tilde{g}_{f_im}}^{*})^{l_i} \nonumber \\
&r=0,\ldots,L, \ \ \ m=1,\ldots,M, \ \ \ k=1,\ldots,K.
\end{align}
The conditions under the  summation symbols ensure that all the terms in \eqref{eq:moments} are distinct.  Even though \eqref{eq:moments} may seem complex, $\mathbb{E}_{|{\bf G}}\{|t_m|^{2r}t_m\varsigma_k^{*}\}$ can be calculated easily for small $r$ values. Note that $r$ in \eqref{eq:C_ysigma_simplified2} is at most $L$, which is typically $1,2,3$, or $4$ when dealing with non-linear hardware \cite{book_rf}, \cite{ericsson}. 

We have now expressed the received signal at the BS in the form ${\bf y}={\bf C}_{y\varsigma}\bm{\varsigma}+\bm{\eta}$ and derived the elements of the effective channel matrix ${\bf C}_{y\varsigma}$. In the following section, we present the LMMSE estimator of the effective channel ${\bf C}_{y\varsigma}$ and develop a deep-learning-based alternative to estimate  ${\bf C}_{y\varsigma}$.

\section{Estimating the Effective Channel}

In this section, we consider channel estimation based on either the standard Bayesian LMMSE methodology or a novel deep learning approach. 
The channel estimation is facilitated by all the users simultaneously sending pilot sequences to the BS.
Let $\tau_{\text{p}}$ denote the uplink training duration in samples per coherence block. Let $\sqrt{\tau_{\text{p}}}\bm{\varphi}_k \in \mathbb{C}^{\tau_{\text{p}}}$ denote the pilot sequence of the $k^\textrm{th}$ UE where $||\bm{\varphi}_k||^2=1$ for $k=1,\ldots,K$. Using the same hardware impairment model as during data transmission, the received baseband signal at the $m^\textrm{th}$ antenna of BS in uplink training phase is given by
\begin{align} \label{eq:baseband-pilots}
{\bf y}^{\text{p}}_{m}={\bf z}^{\text{p}}_m+{\bf n}^{\text{p}}_m, \ \ \ m=1,\ldots,M,
\end{align}
where ${\bf z}^{\text{p}}_m$ is the noise-free distorted signal at the $m^\textrm{th}$ antenna of the BS and ${\bf n}^{\text{p}}_m$ is the uncorrelated thermal noise with ${\bf n}^{\text{p}}_m \sim \mathcal{N}_{\mathbb{C}}({\bf 0}, \sigma^2\bm{I}_{\tau_{\text{p}}})$. The $n^\textrm{th}$ element of ${\bf z}^{\text{p}}_m$ is given by
\begin{align} \label{eq:baseband-pilots2}
z^{\text{p}}_{mn}=&\sum_{l=0}^L\tilde{a}_{lm}\bigg(\sum_{k=1}^Kg_{km}\big(\sqrt{\kappa_kp_k\tau_{\text{p}}}\varphi_{kn}+ \omega_{kn}\big)\bigg) \times \nonumber \\ &\bigg|\sum_{k=1}^Kg_{km}\big(\sqrt{\kappa_kp_k\tau_{\text{p}}}\varphi_{kn}+ \omega_{kn}\big)\bigg|^{2l}, \nonumber \\
&n=1,\ldots,\tau_{\text{p}}, \ \ \ m=1,\ldots,M,
\end{align}
where $\varphi_{kn}$ is the $n^\textrm{th}$ element of $\bm{\varphi}_k$ and $\omega_{kn}$ is a zero-mean circularly symmetric UE distortion term with power $(1-\kappa_k)p_k$ as in Section \rom{2}.B. In \eqref{eq:baseband-pilots2}, $\sum_{k=1}^Kg_{km}\big(\sqrt{\kappa_kp_k\tau_{\text{p}}}\varphi_{kn}+ \omega_{kn}\big)$ is the distortion-free signal at the receiver of the BS without taking into account the BS's hardware impairments. $\{\omega_{kn}\}$ are assumed to be independent of each other, pilot sequences and channels. In the following subsections, we discuss LMMSE and deep-learning-based channel estimation methods.

\subsection{LMMSE Based Channel Estimation}
The MMSE channel estimator is preferred when estimating random variables, but it is hard to compute using the received signal in \eqref{eq:baseband-pilots} since it is not a linear Gaussian model, unlike its distortion-free counterpart in \cite{Kotecha2004a}. 
Previous works have therefore considered the more tractable  Bayesian LMMSE estimation methods \cite{emil_nonideal,emil_book}, which will therefore serve as the benchmark for the deep learning solution we propose. LMMSE estimation can be realized in either a distortion-aware manner, exploiting the second-order moments of the effective channel and additive distortion, or by fully ignoring the impairments. 
 
The conventional approach in MIMO systems is to estimate the physical channels $\{g_{km}\}$. However, the effective channel matrix given in \eqref{eq:C_ysigma_simplified2} is what matters in the data signal detection, thus estimating ${\bf C}_{y\varsigma}$ directly from \eqref{eq:baseband-pilots} is a more effective approach. The distortion-aware LMMSE estimate of the $(m,k)^\textrm{th}$ element of  ${\bf C}_{y\varsigma}$ given ${\bf y}^{\text{p}}_m$ is given by
\begin{align} \label{eq:LMMSE}
[\hat{\bf C}_{y\varsigma}]_{mk}=&{\bf C}_{[C_{y\varsigma}]_{mk}y^{\text{p}}_m}{\bf C}_{y^{\text{p}}_my^{\text{p}}_m}^{-1}{\bf y}^{\text{p}}_m, \nonumber \\ & k=1,\ldots.,K, \ \ \ m=1,\ldots,M,
\end{align}
where ${\bf C}_{[C_{y\varsigma}]_{mk}y^{\text{p}}_m}=\mathbb{E}\{[{\bf C}_{y\varsigma}]_{mk}({\bf y}^{\text{p}}_m)^H\}$ and ${\bf C}_{y^{\text{p}}_my^{\text{p}}_m}=\mathbb{E}\{{\bf y}^{\text{p}}_m({\bf y}^{\text{p}}_m)^H\}$. It is very hard to obtain closed-form expressions for this vector and matrix, thus Monte-Carlo estimation is necessary in the numerical results. Hence, the distortion-aware LMMSE estimator is complicated to use in practice.

The LMMSE estimator in \eqref{eq:LMMSE} effectively treats the distortion as an independent colored noise term, although the distortion is actually a function of the channel. To utilize the inherent distortion structure, we propose a deep-learning-based channel estimation architecture in the next part.

\subsection{Deep Learning Based Channel Estimation}
We propose to use the deep feedforward neural network shown in Fig.~1  in order to realize channel estimation that exploits the structure of the hardware impairments. A feedforward neural network with $P$ fully-connected layers presents a mapping from the input vector ${\bf r}_0\in\mathbb{R}^{N_0}$ to the output vector ${\bf r}_P\in \mathbb{R}^{N_P}$ through $P$ iterative functions:
\begin{align}
{\bf r}_p=\sigma_p({\bf W}_p{\bf r}_{p-1}+{\bf b}_p), \ \ \ p=1,\ldots,P,
\end{align}
where ${\bf W}_p\in \mathbb{R}^{N_p\times N_{p-1}}$ is the weighting matrix at the $p^\textrm{th}$ layer and ${\bf b}_p \in \mathbb{R}^{N_p}$ is the corresponding bias vector. $\sigma_p(\cdot)$ is the activation function for the $p^\textrm{th}$ layer and it is used to introduce non-linearity to the considered mapping. Without this non-linearity, the overall mapping from the input vector to the output vector is simply an affine function. The power of deep learning lies in the use of effective non-linear activation functions in multiple successive ``hidden'' layers. In this way, a properly designed deep neural network can learn how the hardware has impaired the desired signal during uplink training. In particular, it can exploit this information to learn a more effective channel estimation approach compared to the conventional LMMSE method. We refer to \cite{deep_book}, \cite{deep_physical} for further details on deep learning.

Since the large-scale fading coefficients are the same for each antenna of the BS and the small-scale fading coefficients are independent, we will train the neural network in Fig.~1 for a single antenna element and then use it independently for each antenna without loss of generality. Even if the small-scale fading coefficients would be correlated, we can use this structure as a simple and computationally efficient approach. The elements of the effective channels are given in \eqref{eq:C_ysigma_simplified2} and the real and imaginary parts of these are the outputs of the deep neural network in Fig.~1, i.e., $O_k=[{\bf C}_{y\varsigma}]_{mk}$ for $k=1,\ldots,K$. We give the square roots of the large-scale fading coefficients multiplied by the power levels of the UEs, $\{\sqrt{\beta_kp_k}\}$, as input to the neural network in Fig.~1. 
Moreover, we input the processed received pilot signals $\{I_k\}$.
 If we focus on the $m^\textrm{th}$ antenna, these inputs are defined as
\begin{align}
I_k=\bm{\varphi}_k^H{\bf y}_m^{\text{p}}, \ \ \ k=1,\ldots,K,
\end{align}  
where $I_k$ represents the estimate of  $g_{km}$ without taking into account the additional distortion terms in \eqref{eq:baseband-pilots}. Despite its simplicity, the effectiveness of this approach will be shown in the numerical results. 
The ReLU activation function is used in the hidden layers in Fig.~1. This activation function is represented by $[\sigma({\bf u})]_i=\max(0,u_i)$ where the ${\bf u}$ is the input vector of ReLU at each layer.

In training the proposed neural network in Fig.~1, one of the main difficulties is that different users will have different large-scale fading coefficients. We want to train a neural network that can handle any realization of the large-scale fading coefficients (i.e., any set of user locations) without requiring retraining. Hence, we train the network for different realizations and, to simplify the training, we arrange the order of the inputs and outputs such that their indices are according to descending (or ascending) large-scale fading coefficients. We observed empirically that this method works well.
\begin{figure}[t!]
\begin{center}
	\includegraphics[trim={0.1cm 0.05cm 0.1cm 0cm},clip,width=2.5in]{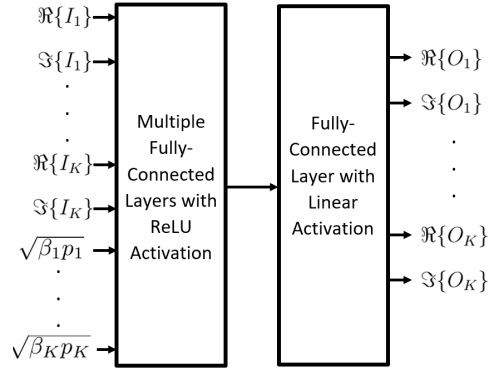} \vspace{-3mm}
	\caption{The proposed deep feedforward neural network for channel estimation with hardware impairments.}
	\label{fig:res}
	\end{center} \vspace{-6mm}
\end{figure}

\vspace{-0.15cm}
\section{Numerical Results}
In this section, we compare the estimation performance of the proposed deep-learning-based channel estimator with two benchmarks: a) 
the LMMSE estimator in \eqref{eq:LMMSE}; and b) the conventional distortion-unaware LMMSE estimator \cite{emil_nonideal}.
The distortion-unaware LMMSE estimator acts as if the BS and UEs have ideal hardware. The polynomial coefficients of the distortion model in \eqref{eq:z_m} are the same for  all the antennas, i.e., $a_{lm}=a_l$ for $m=1,\ldots,M$. Hence, the estimation quality is the same for all antennas and we need not to specify $M$ in the simulations. However, we target a Massive MIMO setup and, therefore, the number of UEs is $K=10$. 

The BS hardware distortion in \eqref{eq:z_m} is modeled as a $2L+1=7$th-order quasi-memoryless polynomial. The polynomial coefficients  are obtained by curve-fitting to the AM/AM and AM/PM distortions of a measured GaN amplifier operating at 2.1 GHz; see \cite{ericsson}. The backoff parameter is selected as $b_{\text{off}}=7$ dB. The $\kappa$-parameter for the UE hardware impairment is the same for all users, $\kappa_k=\kappa$, and the UE distortion terms, $\omega_k$, are modeled as circularly symmetric Gaussian random variables. We assume the QPSK modulation scheme is adopted for uplink data transmission, which determines the effective channel. The pilot length is $\tau_{\text{p}}=K$ and the sequences are the columns of the discrete Fourier transform (DFT) matrix.

There are two hidden layers, each with 300 neurons in the neural network in Fig.~1, and the loss function is the mean squared error (MSE). Both inputs and outputs of the neural network are scaled using the Standard Scaler for $\{I_k, O_k\}$ and MinMax Scaler for $\{\sqrt{\beta_kp_k}\}$. The Adam optimization algorithm is used. The training and validation data lengths are 3,000,000 and 200,000, respectively.   

The signal-to-noise ratio (SNR) of the $k^\textrm{th}$ UE is defined as $\beta_kp_k/\sigma^2$. Fig.~2 shows the normalized MSE (NMSE) of the channel estimates when the SNR is assumed to be equal for all the UEs.  Each point in Fig.~2 presents the average of 100,000 channel realizations. The statistical matrices in the LMMSE estimator in \eqref{eq:LMMSE} are computed using Monte Carlo methods with 100,000 trials. Two different $\kappa$-values are used: 1 and 0.98. Note that the range for practical values of $\kappa$ is between 0.97 and 1 \cite[Sec.~6.1]{emil_book}, where $\kappa=1$ corresponds to perfect UE hardware. Hence, as Fig.~2 shows, the NMSE is larger when having non-ideal UE hardware with $\kappa=0.98$. The distortion-unaware LMMSE and distortion-aware LMMSE in \eqref{eq:LMMSE} perform nearly the same, which is in line with previous results in \cite{emil_nonideal}. As the SNR increases, the proposed deep-learning-based estimator provides substantially lower NMSE. In fact, the proposed method provides an SNR gain of around 3.5 dB and 1.5 dB for $\kappa=1$ and $\kappa=0.98$, respectively, at 20 dB SNR. We conclude that the proposed deep-learning-based estimator captures the structure of the hardware distortion, while the distortion-aware LMMSE estimator fails to do so.

In Fig.~3, the square root of each UE's SNR is chosen randomly from a uniform distribution between $\sqrt{0.1}$ and $\sqrt{100}$ (in linear scale). The cumulative distribution function (CDF) of the NMSE is shown for both the LMMSE and deep-learning-based estimations. The SNRs of the UEs are kept constant through 10,000 small-scale fading channel trials and the statistics required for the LMMSE in \eqref{eq:LMMSE} is obtained by averaging over these trials. Each point in Fig.~3 represents the NMSE obtained over these trials. This is repeated for 1000 different SNR trials resulting in 10,000,000 realizations. We notice that the distortion-aware LMMSE in \eqref{eq:LMMSE} outperforms the distortion-unaware LMMSE only in the low-SNR region (i.e., the upper tail of the CDF curve). The proposed deep-learning-based estimator provides a consistently better estimation quality by exploiting the hardware impairment structure. Similar to Fig.~2, the performance gain  is larger in the case of perfect UE hardware, simply because the UE distortion has no structure to be learned.

\begin{figure}[t!]
	\includegraphics[trim={0.85cm 0.05cm 1.2cm 1.9cm},clip,width=3.58in]{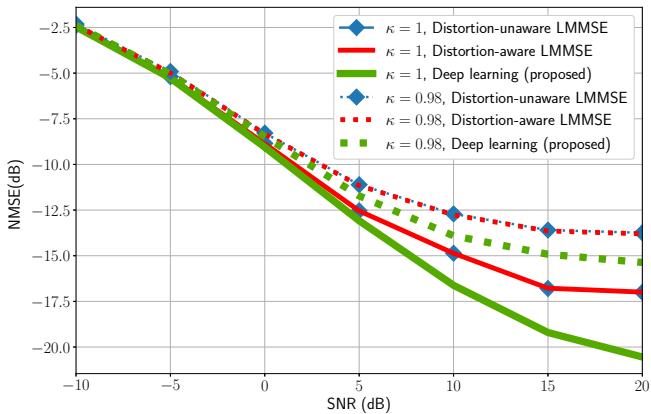} \vspace{-9mm}
	\caption{Normalized mean squared error (NMSE) in dB for equal SNR.}
	\label{fig:sim1} \vspace{-5mm}
\end{figure}

\begin{figure}[t!]
	\includegraphics[trim={1.55cm 0.05cm 0.9cm 1.9cm},clip,width=3.58in]{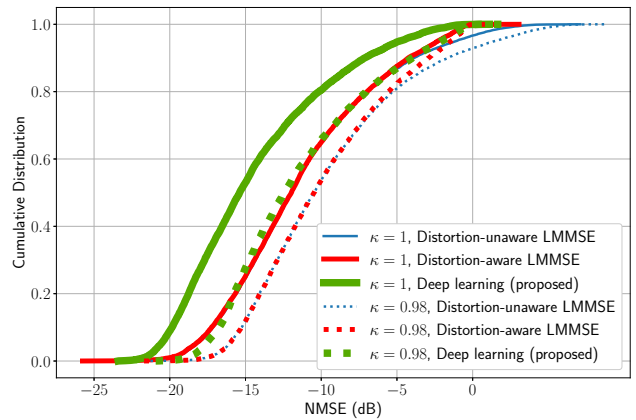} \vspace{-9mm}
	\caption{Normalized mean squared error (NMSE) in dB for varying SNR.}
	\label{fig:sim2} \vspace{-5mm}
\end{figure}
\vspace{-0.15cm}
\section{Conclusions}
\vspace{-0.15cm}
Deep learning has been previously used to learn channel estimators with ideal transceiver hardware \cite{Neumann2018}.
We have proposed a new deep-learning-based channel estimator for systems with hardware impairments, with focus on uplink multi-user MIMO systems. The neural network was trained to utilize the hardware distortion characteristics to achieve better channel estimation quality than with the conventional Bayesian LMMSE estimator, which treats the distortion as an independent colored noise and only utilizes its second-order statistics.
In contrast, the proposed neural network can learn the structure of the quasi-memoryless polynomials that were used for modeling the BS hardware impairments. In the equal-SNR scenario, the performance gap between the proposed deep-learning-based estimator and LMMSE increases with the SNR value. When considering UEs with varying SNRs, the same neural network can be utilized to provide better channel estimates for all the UEs.

Even though the proposed deep learning network is trained using Rayleigh fading, the expressions for effective channels  are valid for any channel model. It can be an interesting future work to evaluate the performance of  proposed method for different channel and hardware impairment models.


\vspace{-0.15cm}
\begin{thebibliography}{00}
	\vspace{-0.15cm}
\bibitem{Kotecha2004a}  J. H. Kotecha and A. M. Sayeed, \textquotedblleft Transmit signal design for optimal estimation of correlated MIMO channels,\textquotedblright\ \emph{\  IEEE Trans. Signal Process.,} vol. 52, no. 2, pp. 546-557, 2004.

\bibitem{Neumann2018} D. Neumann, T. Wiese, and W. Utschick, \textquotedblleft Learning the MMSE channel estimator,\textquotedblright\ \emph{ IEEE Trans. Signal Process.,} vol. 66, no. 11, pp. 2905--2917, 2018.
	
\bibitem{emil_book} E. Bj\"ornson, J. Hoydis, and L. Sanguinetti, \textquotedblleft Massive MIMO networks: Spectral, energy, and hardware efficiency,\textquotedblright\ \emph{Found. Trends Signal Process.,} vol. 11, no. 3-4, pp. 154--655, 2017.

\bibitem{book_rf}  T. Schenk, \emph{\ RF Imperfections in High-Rate Wireless Systems: Impact and
	Digital Compensation.} Dordrecht, The Netherlands: Springer, 2008.	

\bibitem{emil_nonideal} E. Bj\"ornson, J. Hoydis, M. Kountouris, and M. Debbah, \textquotedblleft Massive MIMO systems with non-ideal hardware: Energy efficiency, estimation, and capacity limits,\textquotedblright\ \emph{\  IEEE Trans. Inf. Theory,} vol. 60, no. 11, pp. 7112--7139, 2014.


\bibitem{emil_hardware} E. Bj\"ornson, L. Sanguinetti, and J. Hoydis, \textquotedblleft Hardware distortion correlation has negligible impact on UL massive MIMO spectral efficiency,\textquotedblright\ \emph{\  IEEE Trans. Commun.,} vol. 67, no. 2, pp. 1085--1098, Feb. 2019.

\bibitem{christopher} C. Moll\'en, U. Gustavsson, T. Eriksson and E. G. Larsson, \textquotedblleft Impact of spatial filtering on distortion from low-noise amplifiers in massive MIMO base stations,\textquotedblright\ \emph{ IEEE Trans. Commun.,} vol. 66, no. 12, pp. 6050--6067, 2018.

\bibitem{pa} R. Raich and G. Zhou, \textquotedblleft On the modeling of memory nonlinear effects of
power amplifiers for communication applications,\textquotedblright\ in \emph{\ Proc. IEEE DSP
	Workshop,} Oct. 2002.

\bibitem{ericsson} \emph{Further Elaboration on PA Models for NR,} document 3GPP TSG-RAN
WG4, R4-165901, Ericsson, Aug. 2016.



\bibitem{deep_book} I. Goodfellow, Y. Bengio, and A. Courville, \emph{\ Deep Learning.} Cambridge,
MA, USA: MIT Press, 2016.

\bibitem{deep_physical}  T. O'Shea and J. Hoydis, \textquotedblleft An introduction to deep learning for the physical layer,\textquotedblright \emph{\ IEEE Trans. Cogn. Commun. Netw.,} vol. 3, no. 4, pp. 563--575,
2017.


\end{thebibliography}
\end{document}